\documentclass[preprint,aps,nofootinbib,showpacs]{revtex4}
\usepackage{epsfig}

\newcommand{\as}{\alpha_s}
\newcommand{\BUp}{``Bottom-Up'' }

\begin{document}

\title{Review of the 
``Bottom-Up'' scenario}
\author{Andrej El$^1$
\footnote{E-mail: el@th.physik.uni-frankfurt.de},
Zhe Xu$^1
$\footnote{E-mail: xu@th.physik.uni-frankfurt.de} and 
Carsten Greiner$^1$
\footnote{E-mail: carsten.greiner@th.physik.uni-frankfurt.de}}
\affiliation{$^1$Institut f\"ur Theoretische Physik, Johann Wolfgang 
Goethe-Universit\"at Frankfurt, Max-von-Laue-Str.1, 
D-60438 Frankfurt am Main, Germany}

\begin{abstract}
The thermalization of a longitudinally expanding color glass condensate
with  Bjorken boost invariant geometry is investigated within microscopical
parton cascade BAMPS. Our main focus lies on the detailed comparison of 
thermalization, observed in BAMPS with that suggested in the \BUp scenario.
We demonstrate that the tremendous production of soft gluons via $gg \to ggg$,
which is shown in 
the \BUp picture as the dominant process during the early preequilibration,
will not occur in heavy ion collisions at RHIC and LHC energies,
because the back reaction $ggg\to gg$ hinders the absolute particle
multiplication. Moreover, contrary to the \BUp scenario, soft and
hard gluons thermalize at the same time. The time scale of
thermal equilibration in BAMPS calculations is of order
$\as^{-2} (\ln \as)^{-2} Q_s^{-1}$.
After this time the gluon system exhibits nearly hydrodynamical behavior.
The  shear viscosity to entropy density ratio has a weak
dependence on $Q_s$ and lies close to the lower bound of
the AdS/CFT conjecture.
\end{abstract}
%

\maketitle

\section{Introduction}

It is of great interest to understand the mechanisms of quick equlibration of quark matter at RHIC, indicated by the success of employing simple ideal hydrodynamics \cite{H01} in describing
the large values of the elliptic flow $v_2$ measured in Au+Au collisions \cite{PHENIX,STAR}. 
When the quark gluon
matter becomes sufficiently dilute due to the strong
longitudinal expansion pQCD, bremsstrahlung 
processes are essential for momentum isotropization\cite{XU,XU07}.

The importance of pQCD bremsstrahlung was first raised in the
\BUp scenario \cite{BMSS}, which describes the thermal equilibration
of a Color Glass Condensate (CGC) \cite{CGC,KV} characterized by a saturation scale
$Q_s$. The main idea of the \BUp scenario is that while the hard gluons
with transverse momenta of order $Q_s$ degrade as the condensate evolves in
space time, soft gluons with transverse momenta much smaller than $Q_s$ are
populated due to pQCD $gg\to ggg$ bremsstrahlung and start to dominate the total gluon number. 
As soon as the radiated soft gluons achieve thermal
equilibrium and build up a thermal bath, the hard ones begin to loose
their energy to the thermal bath and subsequently thermalize.
A parametric time scale for overall thermalization is given by
$\tau_{\rm th} \sim \as^{-13/5} Q_s^{-1}$ \cite{BMSS}.

Thermal equilibration of gluonic matter with an idealistic CGC initial condition in presence of pQCD  $gg\leftrightarrow ggg$ processes is investigated in the present work for the first time within transport calculation using the parton
cascade BAMPS \cite{XUG06}. In this work we apply a one dimensional expanding geometry, which is adequate to the assumption
of Bjorken boost invariance \cite{bjorken}. For the initial gluon distribution of CGC we employ an idealized and boost-invariant form \cite{M2000,BV} $f(x,p)=\frac{c}{\as N_c}\frac{1}{\tau}\delta(y-\eta)\Theta(Q_s^2-p_T^2)$, with $N_c=3$ for SU(3) and $c\simeq 0.4$ \cite{KNV,Lappi}. The initial gluons are produced at eigentime $\tau\sim \frac{1}{Q_S}$.  This form of initial condition is a simplified form of CGC inspired by earlier studies in \cite{M2000,BV}. Particles with transverse momenta $p_\perp > Q_s$ are missing here. Their presence would modify the thermalization process, but as well lead to an even faster thermalization of the hard momentum region. Including a more realistic distribution in the hard region of transverse spectrum is a task for future work.  

\section{Results: Thermalization of a CGC}

\begin{figure*}
\hskip -0.5cm
\begin{minipage}[t]{80mm}
\epsfxsize=8.0 cm
\epsfysize=4.0 cm
\epsfbox{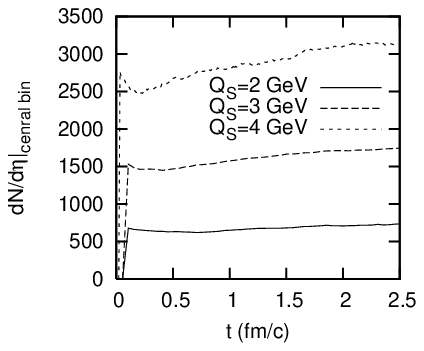}
\vskip -0.25cm
\caption{$\frac{dN}{d\eta}$(central $\eta$ bin),$\alpha_s=0.3$}
\label{dNdeta}
\end{minipage}
\begin{minipage}[t]{80mm}
\epsfxsize=8.0 cm
\epsfysize=4.0 cm
\epsfbox{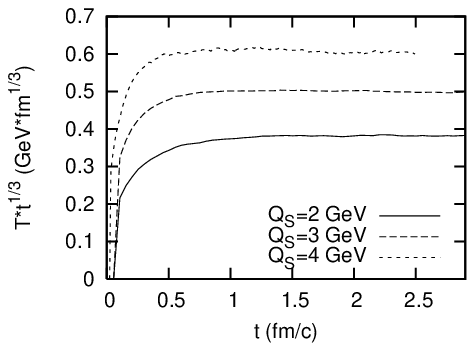}
\vskip -0.25cm
\caption{Scaled effective temperature}
\label{temp}
\end{minipage}
\end{figure*}

The way that thermalization proceeds within the parton cascade calculations
does not resemble the \BUp scenario \cite{BMSS}. The strong parametric enhancement of the total 
gluon number at early times, as predicted by \BUp scenario,
is not observed in the cascade calculations. Instead, gluon annihilation 
occurs during the first $0.3-0.75$ fm/c for $Q_s=2-4$ GeV, as shown in Fig. \ref{dNdeta}, which is due to domination of the $3\to 2$ processes at the early stage of evolution. 
This indicates that the initial CGC is oversaturated for the chosen values of $\as$ and $Q_s$.  The gluon number begins to increase after $t\sim 0.3-0.75$ fm/c (depending on the value of $Q_s$) when the system is close to kinetic
equilibrium and a quasi-hydrodynamical cooling sets in.  

\begin{figure*}
\hskip -0.5cm
\begin{minipage}[t]{100mm}
\epsfxsize=11.0 cm
\epsfysize=4.5 cm
\epsfbox{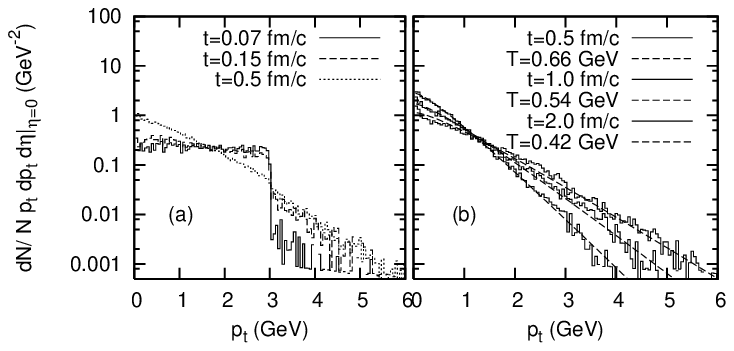}
\vskip -0.3cm
\caption{  Transverse momentum spectra at early times (a) and  for later times (b)}
\label{nnormpt_early}
\end{minipage}
\begin{minipage}[t]{60mm}
\epsfxsize=5.0 cm
\epsfysize=4.5 cm
\epsfbox{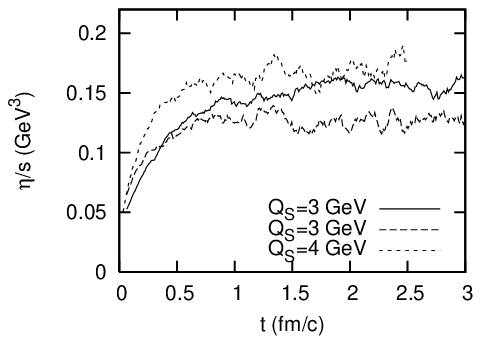}
\vskip -0.3cm
\caption{Ratio of the shear viscosity to the entropy density.}
\label{etaovers}
\end{minipage}
\end{figure*}

After the expansion starts, energy flows immediately into both the soft
($p_t < p_{\rm soft}=1.5$ GeV) and hard momentum region
($p_t > Q_s$) where the populations rapidly increase, as demonstrated in Fig.\ref{nnormpt_early}(a) for calculations with $\as=0.3$ and $Q_s=3$ GeV. The presence of a thermal bath 
of soft gluons seems not to be a necessary condition for the equilibration
of hard gluons, since they achieve an exponential shape on a short time scale and almost as quick as
the soft gluons, as Fig.\ref{nnormpt_early}(a) demonstrates. Again, this is different from the picture invoked in
the \BUp scenario. 

At $t=0.5 fm/c$ the transverse spectrum  looks almost thermal.
After $t\approx 0.5$ fm/c  the cooling of the system sets in and the spectra evolve in a way, characteristic for a hydrodynamical expansion, as shown in Fig.\ref{nnormpt_early}(b) for $\as=0.3,Q_s=3~GeV$.

We extract the time scale when the system is more or less thermalized from Fig. \ref{temp}, 
where the scaled effective temperature $T t^{1/3}$ is shown. We define the time scale
of thermalization  as the time, at which $T t^{1/3}$ becomes a constant, which corresponds to a one-dimensional ideal
hydrodynamical expansion. The times extracted from Fig. \ref{temp}(constant $\alpha_s=0.3$) read: $1.2,~0.75,~0.55 fm/c$ (resp. for $Q_s=2,3,4~GeV$). For fixed $Q_s=3$ GeV and various $\as$ the thermalization times are extracted in same manner:
$1.75,~1.0,~0.75~fm/c$ (resp. for $\as=0.1,0.2,0.3$).

From the values presented here we observe that the time scale of thermalization is of order 
$\as^{-2} (\ln \as)^{-2} Q_s^{-1}$. The $1/Q_s$ scaling is consistent with the \BUp result. The dependence on $\alpha_s$  proves to be weaker than the \BUp prediction but consistent with the scaling obtained in \cite{XGviscos}. The reason for the quick thermalization observed in our calculations is the gluon bremsstrahlung, which favors large-angle radiation due to the LPM suppression\cite{XU,XU07,XUG06}. Thus, the pQCD $gg\leftrightarrow ggg$
processes are dominant for the thermal equilibration.

The ideal hydrodynamic behavior of the system at late times indicates that  the ratio of the shear viscosity to the
entropy density is small. Figure \ref{etaovers} shows the $\eta/s$ ratio in the calculations
with $\as=0.3$ and $Q_s=2$, $3$, and $4$, respectively. The shear viscosity is calculated using the Navier-Stokes approximation: $\eta=\frac{t}{4}\left(T_{xx}+T_{yy}-2 T_{zz}\right)$. The entropy density is calculated by $s=4n-n\, \ln (\lambda)$ , where $n$ is the gluon density and $\lambda=n/n_{\rm th}$ denotes
the gluon fugacity.

The $\eta/s$ ratio is nearly constant and has a weak dependence
on $Q_s$, $\eta/ s \approx 0.15$, which is exactly the same as that 
obtained in full 3+1 dimensional BAMPS calculations with $\as=0.3$
and minijets type initial conditions for Au+Au collisions at RHIC
energies \cite{XGSv2}. This verifies that the $\eta/s$ ratio determines
the behavior of the late dynamics and, thus, depends only on the coupling
$\as$, but not on initial conditions. The smallness of the $\eta/s$ ratio, which is close to the AdS/CFT lower bound\cite{ads}, corresponds to the efficiency of the pQCD $gg\leftrightarrow ggg$ processes in
thermal equilibration, because the $\eta/s$ ratio is inversely proportional
to the total transport collision rate which is for $gg\leftrightarrow ggg$ processes $6-7$ times larger
than that of $gg\to gg$ collisions \cite{XGviscos}.

\section{Conclusion}
\label{con}
 Using the parton cascade BAMPS, we found that several
aspects of the real thermalization might be different compared to
the \BUp scenario. A strong increase in soft gluon number, as
predicted in the \BUp scenario, is not observed and thermal equilibration of soft and hard gluons occurs
roughly on the same time scale due to $2\to 3$ and $3\to 2$ processes,
respectively. In agreement with the \BUp scenario, the thermalization
time proves to be proportional to $Q_s^{-1}$, however, its proportionality
to $\as^{-13/5}$ is not seen but is much weaker: 
$\tau_{\rm th}\sim (\as \ln \as )^{-2} Q_s^{-1}$. The differences arise because the back reactions of bremsstrahlung, $3\to 2$ processes, play a significant role. They are completely absent in the \BUp scenario. 

The quick thermalization and the smallness of the $\eta/s$ ratio observed
in the present calculations with the CGC initial conditions are
consistent with the findings from the previous 
studies \cite{XU,XU07,XUG06,XGviscos,XGSv2}
using the Glauber-type minijets initial conditions. This demonstrates that
independent of the chosen initial conditions,
the pQCD bremsstrahlung processes (and the back reactions) dominate
the dynamical equilibration and then keep the system behaving like
a nearly perfect fluid.

\end{document}